\documentclass[sigconf, nonacm]{acmart}

\newcommand\vldbpagestyle{plain} 

\usepackage{tikz}
\usepackage{multirow}
\usepackage{makecell}
\usepackage{filecontents}
\usepackage{booktabs}
\usepackage{amsmath}
\usepackage{amsfonts}
\usepackage{pifont} 
\usepackage{epstopdf} 
\usepackage{enumitem}

\usepackage[compact]{titlesec}
\titleformat*{\section}{\large\bfseries}
\titleformat*{\subsection}{\normalsize\bfseries}
\titleformat*{\subsubsection}{\normalsize\bfseries}
\titlespacing*{\section}{0pt}{*2}{*1}
\titlespacing*{\subsection}{0pt}{*1.5}{*0.8}
\DeclareTextFontCommand{\textbfit}{\fontseries\bfdefault \itshape
}

\usepackage{epsfig,makecell}
\usepackage{subfigure}
\usepackage[labelfont=bf,font=normal,skip=1pt]{caption}

\usepackage{algorithm}
\usepackage{algpseudocode}

\usepackage{tikz}

\newcommand{\sys}{\textsc{ProMoE}}

\newcommand{\diffatomic}[1]{#1}

\newcommand{\ra}[1]{\renewcommand{\arraystretch}{#1}}

\newcommand{\eat}[1]{}
\newcommand{\done}[1]{}

\newcommand{\nospacestitle}[1]{\noindent{\bf #1}}
\newcommand{\stitle}[1]{\vspace{1.ex}\noindent{\bf #1}}

\newcommand{\sstab}{\rule{0pt}{8pt}\\[0.5ex]}

\newcommand{\pmetric}{\textsf{GoodPred}}
\newcommand{\accuracy}{\textsf{Accuracy}}
\newcommand{\early}{\textsf{FetchRate}}

\begin{document}

\title{{\sys}: Fast MoE-based LLM Serving using Proactive Caching}

\author{Xiaoniu Song$^{1}$ \ \ Zihang Zhong$^{3,*}$ \ \ Rong Chen$^{1,\ddagger}$ \ \ Haibo Chen$^{1,2}$} \thanks{$^{\ddagger}$ Rong Chen is the corresponding author (\url{rongchen@sjtu.edu.cn}).}
\thanks{$^{*}$ During internship at Shanghai Jiao Tong University.}

\affiliation{\vspace{1.5ex}
\institution{$^1$Institute of Parallel and Distributed Systems, Shanghai Jiao Tong University}
\country{}
}
\affiliation{\vspace{.5ex}
\institution{$^2$Key Laboratory of System Software (Chinese Academy of Sciences) \ \ $^3$Zhejiang University}
\country{}
\vspace{1.5ex}
}
\renewcommand{\shortauthors}{X. Song, Z. Zhong, R. Chen, and H. Chen}

\begin{abstract}

The promising applications of large language models are often limited by the constrained GPU memory capacity available on edge devices. 
Mixture-of-Experts (MoE) models help address this issue by activating only a subset of the model's parameters during computation.
This approach allows the unused parameters to be offloaded to host memory, thereby reducing the overall GPU memory demand.
However, existing cache-based offloading solutions handle cache misses reactively, which significantly impacts system performance.
In this paper, we introduce {\sys}, a novel proactive caching system that utilizes intermediate results to predict subsequent expert usage.
By proactively fetching experts in advance, {\sys} eliminates passive cache misses, removes loading time from the critical path, and reduces the performance overhead associated with offloading.
Our evaluations demonstrate that {\sys} achieves an average speedup of 2.20$\times$ (up to 3.21$\times$) and 2.07$\times$ (up to 5.02$\times$) in the prefill and decode stages, respectively, compared to existing offloading solutions.

\end{abstract} 
\maketitle

\pagestyle{\vldbpagestyle}

\section{Introduction}
\label{sec:intro}

Large language models (LLMs) have revolutionized various fields, 
including natural language processing, content generation, 
and decision support~\cite{brown2020language, zhang2022opt, touvron2023llama,rlhf,kaplan_scaling_2020}. 
Traditionally, these models have been deployed in data centers equipped with high-end GPUs. 
However, there is growing interest in running LLMs on consumer-grade platforms 
to enhance privacy and speed~\cite{edge-llm-2,edgemoe,song_powerinfer_2023}. 
Despite this growing interest, significant challenges remain 
due to memory limitations. LLMs typically require substantial memory 
(often hundreds of gigabytes)~\cite{touvron2023llama,zhang2022opt,kaplan_scaling_2020}, 
which exceeds the capacities of consumer-grade GPUs, generally limited to around a dozen gigabytes. 
This limitation leads to serious performance issues, 
ultimately hindering the efficiency and adoption of LLMs on personal computers.

Mixture-of-Experts (MoE)~\cite{jacobs_adaptive_1991,mixtral,yang_qwen2_2024,dai_deepseekmoe_2024,deepseek-ai_deepseek-v2_2024}
offers an opportunity to address the GPU memory constraints 
faced by LLMs by dividing the model into multiple experts and activating only a few during inference. 
This approach allows most experts to be offloaded to host memory, loading only the necessary ones 
into GPU memory. 
While this significantly reduces GPU memory requirements, 
\emph{expert offloading} also introduces severe performance degradation up to 8.9$\times$~\cite{kong2024swapmoe}
due to limited PCIe bandwidth between host and GPU memory (32GB/s unidirectional on PCIe 4.0).

\begin{figure}[t]
\begin{minipage}{1\linewidth}
  \centering
  \includegraphics[width=1\textwidth]{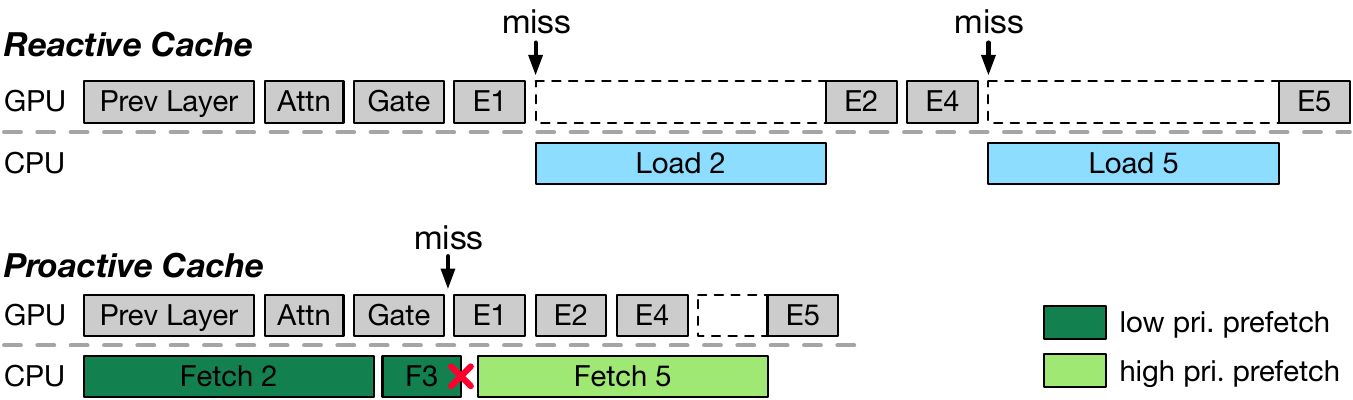}
  \end{minipage} \\[4pt]
  \begin{minipage}{1\linewidth}
  \caption{\small{\emph{A comparison of execution flow between reactive and proactive caching.
  }}}
  \label{fig:intro:proactive-cache}
  \end{minipage} \\[-15pt]
\end{figure}

Recently, researchers have proposed caching frequently accessed experts in GPU memory 
to minimize offloading costs~\cite{mixtral_offloading}.
However, this caching approach handles missing experts in a \textbf{reactive} manner. Specifically,
a cache miss is triggered passively when an expert is accessed during inference, leaving the expensive loading 
on the critical path (see Figure~\ref{fig:intro:proactive-cache}).
For instance, when caching 50\% of the experts in the deepseek-moe~\cite{ds1} model, the time spent on loading missing experts accounts for over 60\% of the total inference time.
Additionally, the inherent low skewness and poor locality of expert access patterns in MoE models, especially in modern decoder-only architectures, significantly limit 
the potential improvements that can be achieved through better caching policies.

\begin{figure*}[t!]
\begin{minipage}{1\linewidth}
\centering
\includegraphics[width=1\linewidth]{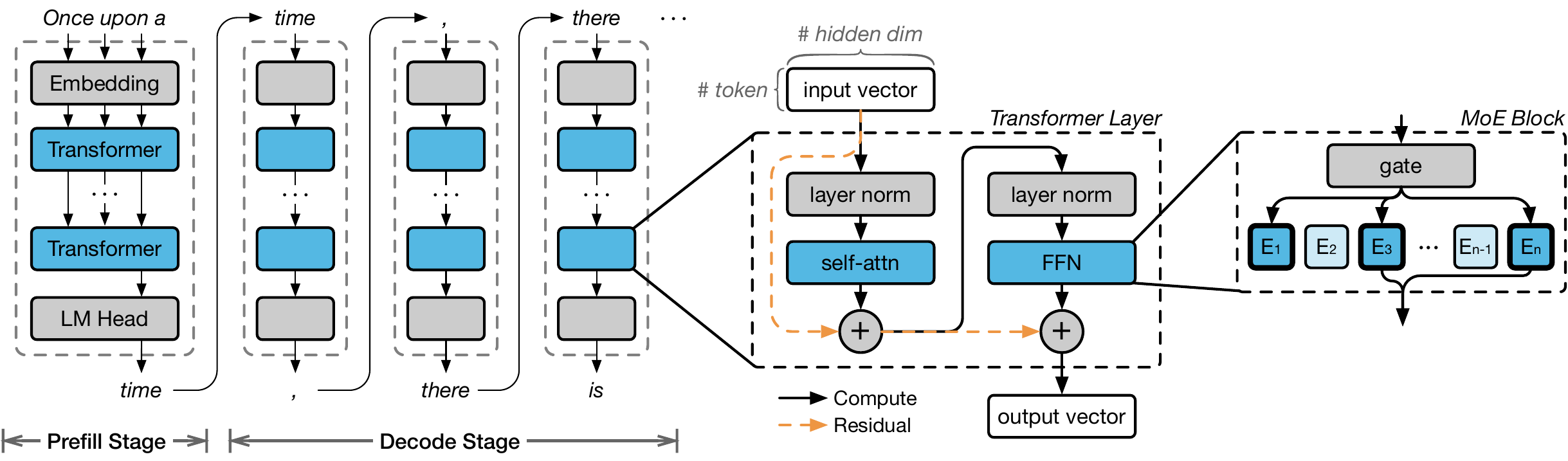}
\end{minipage} \\[4pt]
\begin{minipage}{1\linewidth}
\caption{\small{\emph{
(a) The execution flow of large language models (LLMs),
(b) the architecture of a transformer layer in LLMs, and
(c) the architecture of a Mixture-of-Experts (MoE) block that replaces FFN in a transformer layer.
}}}
\label{fig:bg:llm-arch-in-one}
\end{minipage} \\[-5pt]
\end{figure*}

In this paper, we propose {\sys}, a novel system to address the performance challenges associated with offloading in MoE-based LLMs through \textbf{proactive} caching, as shown in Figure~\ref{fig:intro:proactive-cache}.
By actively predicting which experts will be needed and prefetching their parameters into a cache in GPU memory,
{\sys} can take the time required for fetching missing experts off the critical path.
This allows for better overlap with computation, enhancing overall performance and GPU utilization.

To achieve effective proactive caching, {\sys} addresses two main questions.
First, given the dynamic nature of MoE models, {\sys} requires a predictive approach for prefetching.
To evaluate the quality of a prediction method, {\sys} introduces a metric called {\pmetric}.
This metric considers both the accuracy and efficiency of the predictions.
To achieve a high {\pmetric} score, {\sys} proposes a learned predictor 
that prefetches experts in a stride manner. This learned predictor identifies correlations between intermediate results and expert selections
, allowing for accurate predictions of experts while the stride prefetching technique 
perfectly hides prediction latency, ensuring high efficiency of prefetching.

Second, the processes of prefetching and inference can interfere with each other, 
leading to low utilization of the GPU, cache, and bandwidth for prefetching.
Therefore, {\sys} needs to carefully coordinate these two processes to minimize interference.
We observed that the required experts for each layer can be identified all at once, 
which creates opportunities to optimize prefetching and inference for better overlap.
Based on this insight, {\sys} proposes several techniques to coordinate 
the execution of prefetching and inference processes, 
including chunked prefetching, early preemption, and reordered inference.
These techniques eliminate passive cache misses and maximize the overlap 
between prefetching and inference, thereby reducing inference latency and improving utilization.

To demonstrate the effectiveness of {\sys} in serving MoE-based LLMs on consumer-grade hardware, we integrated {\sys} into two widely used LLM frameworks: transformers and llama.cpp.
Compared to hand-crafted caching baselines with state-of-the-art performance, {\sys} achieves an average speedup of 1.78$\times$ (up to 2.48$\times$) in the prefill stage and 1.34$\times$ (up to 1.79$\times$) in the decode stage.
When compared to existing offloading methods available in open-source LLM frameworks, {\sys} achieves an average speedup of 2.20$\times$ (up to 3.21$\times$) and 2.07$\times$ (up to 5.02$\times$) for these two stages.
The source code of {\sys} is publicly available at {{\url{https://github.com/promoe-opensource/promoe}}}.

\stitle{Contributions}. We make the following contributions.
\sstab (1) A new metric called {\emph{``\pmetric''}} that can holistically 
evaluate various predictors in expert prefetching ($\S$\ref{sec:predict}).\sstab (2) A novel learned predictor, coupled with a stride mechanism, 
that achieves high accuracy while hiding prediction latency ($\S$\ref{sec:predict}).\sstab (3) A sophisticated proactive cache that eliminates passive cache misses by coordinating prefetching and inference ($\S$\ref{sec:prefetch}).\sstab (4) An implementation integrated into mainstream LLM frameworks ($\S$\ref{sec:impl}), 
along with an evaluation showing the efficacy and efficiency of {\sys} 
compared to state-of-the-art solutions ($\S$\ref{sec:eval}).

\section{Background}
\label{sec:bg}

\begin{table}[t]
\vspace{2pt}
\begin{minipage}{1.0\linewidth}
\caption{\small{\emph{MoE-based LLMs description. 
\textup{\textbf{P}, \textbf{L}, and \textbf{E}} denote parameters, layers, and experts, respectively.
\textup{\textbf{Act.}} indicates the number of activated parameters or experts 
during the inference of a single token.
}}}
\label{tbl:moe-models}
\end{minipage} \\[2pt]
\begin{minipage}{1.0\linewidth}
\ra{1.1}
\centering
\begin{tabular}{@{~}lrrrrr@{~}}
\toprule

\multirow{2}{*}{\textbf{MoE-based LLM}}
& \multicolumn{2}{c}{\textbf{\#P}}
& \multirow{2}{*}{\textbf{\#L}}
& \multicolumn{2}{c}{\textbf{\#E per L}} \\

\cmidrule(lr){2-3}
\cmidrule(lr){5-6}

& \textbf{Total} & \textbf{Act.} & & \textbf{Total} & \textbf{Act.} \\

\midrule
Deepseek-moe (DS-1)~\cite{ds1}      & 16.4B &  2.8B & 28 & 64 & 6 \\
Deepseek-v2-lite (DS-2)~\cite{ds2}  & 15.7B &  2.7B & 27 & 64 & 6 \\
Qwen1.5-moe (QW-1)~\cite{qw1}       & 14.3B &  2.7B & 24 & 60 & 4 \\
Qwen2-moe (QW-2)~\cite{qw2}         & 57.4B & 14.2B & 28 & 64 & 8 \\
Mixtral-8x7B (Mixt)~\cite{mixt}     & 46.7B & 12.9B & 32 &  8 & 2 \\
\bottomrule
\end{tabular}
\end{minipage} \\[-10pt]
\end{table}

\subsection{Mixture-of-Experts (MoE) based LLMs}

Large Language Models (LLMs) perform inference in two stages: prefill and decode, as illustrated in Figure~\ref{fig:bg:llm-arch-in-one}(a).
During the prefill stage, the model processes the user's input prompt in a single iteration.
The tokens within the prompt are processed in parallel by the model, and the first token of the response is generated at the end of this iteration.
In the decode stage, each iteration processes only one token generated from the previous iteration, producing the next token.
These tokens are fed into the model sequentially and ultimately concatenated to form the complete response.
Due to the differing computational scales between the two stages of LLM inference, their performance is typically measured separately.
The performance of the prefill stage is usually quantified by the Time to First Token (TTFT), which represents the duration users wait for the LLM to process the prompt before it starts generating output.
For the decode stage, performance is commonly measured using either Tokens Per Second (TPS) or Time Per Output Token (TPOT).

Large Language Models (LLMs) consist of a series of transformer layers.
Each layer contains a self-attention block (self-attn) and a feed-forward network (FFN), 
as shown in Figure~\ref{fig:bg:llm-arch-in-one}(b).
These components process the input hidden states, add the results back to the inputs, and pass them to the next layer.
Due to layer normalization, the outputs are numerically smaller than their inputs, 
leading to a slow change in hidden states across layers~\cite{lee_infinigen_2024,liu_deja_2023}.
Typically, the cosine similarity between hidden states of adjacent layers averages around 90\%.

The Mixture-of-Experts (MoE) architecture enhances LLMs by expanding the FFN into multiple experts,
as depicted in Figure~\ref{fig:bg:llm-arch-in-one}(c).
This approach increases the model's parameters while reducing overall computation, 
since only a subset of experts is activated during each forward pass.
Specifically, each MoE block consists of a gate function and multiple experts.
The gate function prioritizes which experts should process the current token.
Each expert is structurally similar to the original FFN but contains fewer parameters.
The output of the MoE block is a weighted average of the outputs from all activated experts.

In MoE-based LLMs, expert selection occurs independently for each token, as shown in Table~\ref{tbl:moe-models}.
When processing multiple tokens simultaneously (e.g., when processing prompts
or batching multiple requests), a larger portion of experts is activated, ranging from over 50\% to nearly 100\%, depending on the number of tokens.

\diffatomic{
\subsection{Caching MoE-based LLMs}

\begin{figure}[t]
\vspace{2mm}
\begin{minipage}[t]{1\linewidth}
\centering
\includegraphics[width=0.95\linewidth]{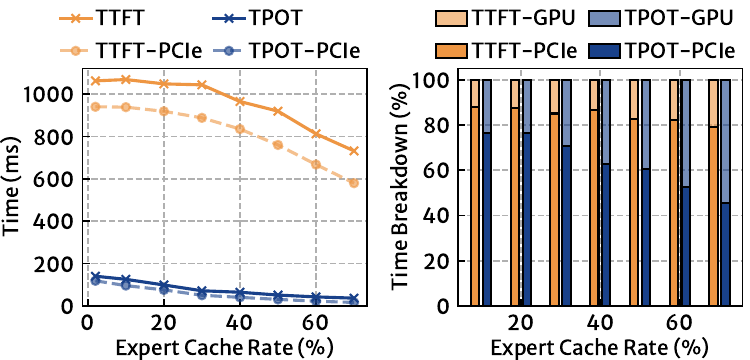}
\end{minipage} \\ [4pt]
\begin{minipage}{1\linewidth}
\caption{\small{\emph{
The (a) latency and (b) time breakdown of LRU caching under different cache rates in transformers with DS-1 model.
}}}
\label{fig:motiv:latency-of-caching-trans}
\end{minipage}  \\[-5pt]
\end{figure}
}

\diffatomic{
\begin{figure}[t]
\vspace{1mm}
\begin{minipage}[t]{1\linewidth}
\centering
\includegraphics[width=0.95\linewidth]{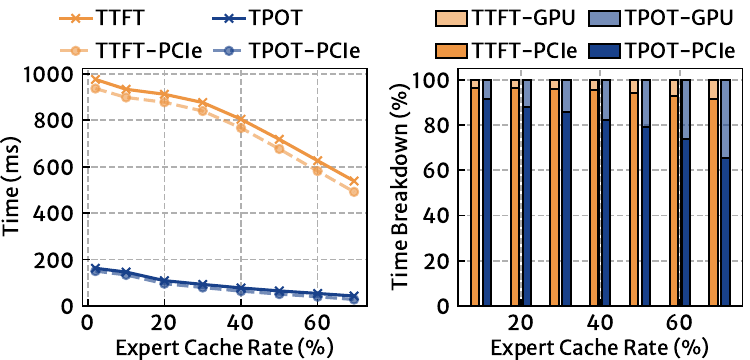}
\end{minipage} \\ [4pt]
\begin{minipage}{1\linewidth}
\caption{\small{\emph{
The (a) latency and (b) time breakdown of LRU caching under different cache rates in llama.cpp with QW-2 model.
}}}
\label{fig:motiv:latency-of-caching-llama}
\end{minipage}  \\[-10pt]
\end{figure}
}

In MoE-based LLMs, each token utilizes only a subset of experts.
Most experts can be offloaded to CPU memory, with only the necessary experts loaded into GPU memory.
This allows MoE-based LLMs to run on consumer-grade hardware with limited GPU memory.
However, due to the limited PCIe bandwidth, directly offloading parameters to CPU memory can lead to high latency and low GPU utilization.
For instance, when running the DS-1 model with 50\% of experts offloaded to CPU memory, the TPOT is 67.9 ms, while fetching experts from host memory takes 58.1 ms, accounting for 85.6\% of the total time.
Each output token requires 2.67 GiB of expert parameters in FP16 precision, with 1.33 GiB needing to be transferred from CPU memory to GPU memory due to offloading.
The achieved bandwidth is 23 GB/s, which matches the achievable bandwidth (23.9 GB/s in our bandwidth test) from host to GPU using PCIe 4.0x8.

To mitigate the performance issues caused by offloading, a traditional method is to cache frequently accessed experts in GPU memory.
A common approach is to use LRU (Least Recently Used) or static caching to store these frequently accessed experts.
For example, Mixtral-offloading~\cite{mixtral_offloading} implements an LRU cache for the Mixtral model.
Another example is CUDA's Unified Memory (UM), which leverages a paging mechanism to transfer data between the GPU and CPU on demand.

\begin{figure}[t]
\begin{minipage}[t]{1\linewidth}
\centering
\includegraphics[width=1\linewidth]{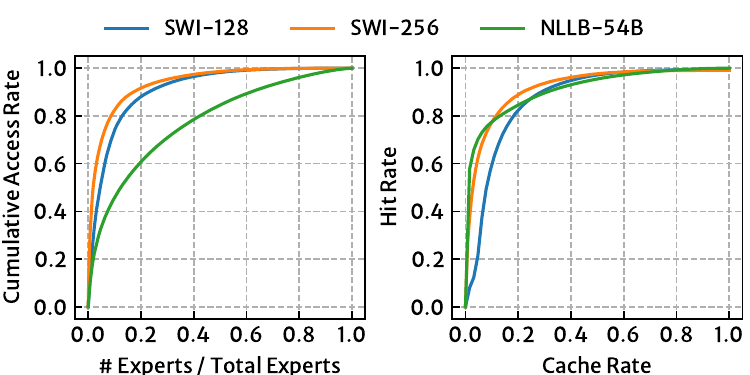}
\end{minipage} \\ [2pt]
\begin{minipage}[t]{1\linewidth}
\centering
\includegraphics[width=1\linewidth]{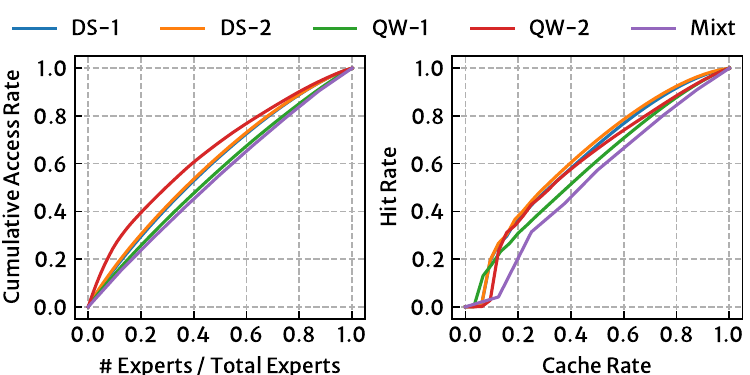}
\end{minipage} \\ [4pt]
\begin{minipage}{1\linewidth}
\caption{\small{\emph{
The (a) CDF of expert access frequency and (b) hit rate of LRU caching.
The upper figures show results of traditional encoder-decoder models (switch-transformer~\cite{fedus2022switch}, NLLB~\cite{costa2022no}), while the lower figures show results of modern decoder-only models listed in Table~\ref{tbl:moe-models}.
}}}
\label{fig:motiv:cache-hit-rate}
\end{minipage}  \\[-10pt]
\end{figure}

The major challenge of caching in MoE is its \textbf{reactive} nature when handling cache misses.
When the inference process encounters an expert that is missing from GPU memory, the computation is blocked until the expert is fetched from host memory,
resulting in high latency overhead in the critical path of inference.

We evaluated the performance of LRU caching in transformers~\cite{transformers} with the DS-1 (fp16) model, along with llama.cpp~\cite{llamacpp} using the QW-2 (int4) model.
Figure~\ref{fig:motiv:latency-of-caching-trans} and~\ref{fig:motiv:latency-of-caching-llama} illustrate the inference latency and the time breakdown for both the prefill and decode stages.
In the case of the DS-1 model, caching 50\% of experts results in a blocking time of 60.4\% on the critical path during the decode stage.
The blocking time during the prefill stage is more severe, as more experts are accessed, leading to an 82.7\% blocking time on the critical path.
For llama.cpp, which achieves faster inference by eliminating the overhead of the Python interpreter, the proportion of blocking time is even greater.
Caching 50\% of experts results in 94.2\% blocking time during the prefill stage and 79.0\% during the decode stage.

Another factor that exacerbates the impact of blocking time on the critical path is the access frequencies of different experts in MoE-based LLMs, particularly modern decoder-only LLMs, which tend to be less skewed.
Figure~\ref{fig:motiv:cache-hit-rate} shows the cumulative access frequency of experts and the hit rate of LRU caching.
Traditional encoder-decoder MoE models like switch-transformer~\cite{fedus2022switch} (SWI) and NLLB~\cite{costa2022no} released in 2022 exhibit a power-law distribution in expert access frequencies, where a small number of experts are accessed more frequently than others.
This high skewness leads to high hit rates and benefits both static and dynamic caching like LRU.
In contrast, modern decoder-only MoE models exhibit a more uniform access pattern, as shown in the bottom of Figure~\ref{fig:motiv:cache-hit-rate}.
This low skewness creates unique challenges for offloading and caching in modern MoE models.

This uniform access pattern can be attributed to the deliberate design of modern MoE models, which utilize various techniques during training to prevent any single expert from becoming a hotspot.
This is crucial because uneven expert utilization can lead to inadequate training of certain experts, ultimately impacting the model's performance.
This phenomenon is referred to as ``routing collapse''~\cite{routing-collapse}.
To mitigate routing collapse, contemporary MoE models incorporate strategies such as Device-Limited Routing~\cite{deepseek-ai_deepseek-v2_2024} and Expert-Level Balance Loss~\cite{dai_deepseekmoe_2024} during the training process.
Consequently, the access frequencies of different experts tend to be more uniform during inference.
Combined with the reactive handling of cache misses, the caching solution significantly degrades the critical path latency. 
\begin{figure}[t]
  \vspace{2mm}
  \begin{minipage}{.98\linewidth}
  \centering
\includegraphics[width=1\linewidth]{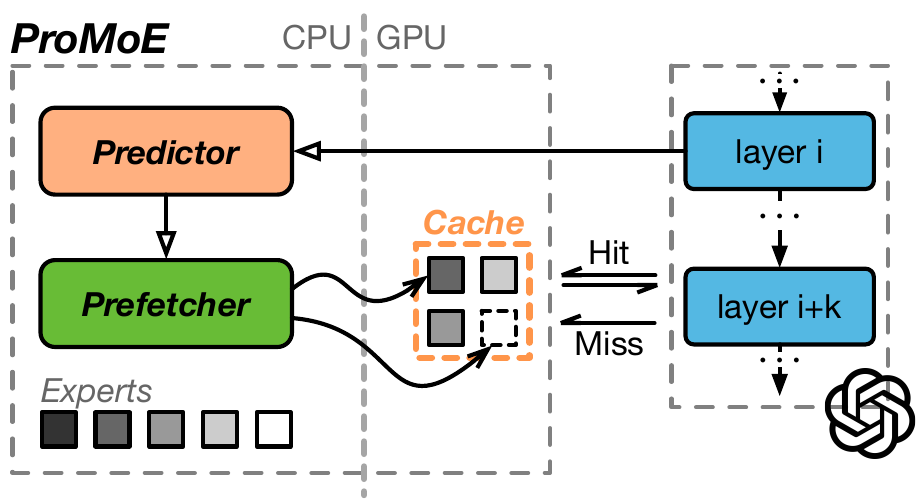}
  \end{minipage} \\[4pt]
  \begin{minipage}{1\linewidth}
  \caption{\small{\emph{The architecture of {\sys}.
  }}}
  \label{fig:overview:sys-arch}
  \end{minipage} \\[-15pt]
\end{figure}

\section{Overview of {\sys}}

This paper presents {\sys}, a system that achieves low-latency inference for MoE-based LLMs on consumer-grade platforms.
{\sys} addresses the reactive nature of existing solutions, which passively trigger data transfers on the critical path of inference, leading to high latency.
To tackle this issue, {\sys} adopts a proactive caching approach.
Instead of directly reducing data transfers between the CPU and GPU,
proactive caching moves data transfers out of the critical path, allowing them to overlap with inference.

The architecture of {\sys} is illustrated in Figure~\ref{fig:overview:sys-arch}.
It consists of two main components: the predictor and the prefetcher.
The predictor periodically predicts which experts will be selected.
Based on these predictions, the prefetcher preloads experts into the GPU cache.
During inference, the LLM inference engine accesses experts stored in the cache and triggers misses for any experts that are absent.
Compared to existing solutions, most expert data transfers in {\sys} occur outside the critical path of inference, thus reducing latency and improving GPU utilization.

To achieve effective proactive caching, {\sys} must address the questions of ``\textit{what to prefetch}'' and ``\textit{how to prefetch}'' as mentioned in $\S$\ref{sec:intro}.
The predictor in {\sys} tackles the first question by making good predictions.
To define what constitutes a good prediction, {\sys} proposes a {\pmetric} metric that considers both the accuracy and efficiency
of the predictions.
Based on this metric, {\sys} introduces a learned predictor that prefetches experts in a stride manner.
This learned predictor memorizes the correlations between intermediate results and expert selections to make accurate predictions of expert selections.
Additionally, through stride prefetching, {\sys} overlaps the processes of prediction and prefetching to 
hide the latency of predictor.

{\sys}'s prefetcher addresses the second question by carefully coordinating the prefetching and inference processes.
Naive prefetching can lead to interference between these processes, resulting in suboptimal performance.
{\sys} leverages the observation that the choice of experts for each layer becomes available all at once after the gating function.
Based on this insight, {\sys} proposes three key techniques to effectively coordinate prefetching and inference: chunked prefetching, early preemption, and reordered inference.
With these techniques working in concert, {\sys} can eliminate passive cache misses and maximize the overlap between prefetching and inference, thereby reducing inference latency.
 
\section{GoodPred, Prediction, and Prefetching}
\label{sec:predict}

The dynamic nature of MoE models necessitates the deployment of a predictor in {\sys} 
to make approximate predictions of experts for prefetching.
To ensure effective prefetching, the predictor must meet two primary requirements: 
accuracy and efficiency. In this section, we first define a key metric called {\pmetric}, which combines these two aspects 
to evaluate the performance of a predictor.
Subsequently, we introduce {\sys}'s learned predictor and explain how it improves both accuracy and efficiency.

\subsection{A New Prediction Metric: GoodPred}
A good predictor requires both high accuracy and efficiency.
Higher accuracy increases the likelihood that predicted experts will be utilized,
while higher efficiency allows more time to load these predicted experts.
These two goals must be pursued simultaneously, though these two goals might initially seem contradictory---improving accuracy 
often requires more prediction time, which can reduce the time available for prefetching.

To assess the performance of the predictor, we define the {\pmetric} metric as follows:
$$
{\pmetric} = {\accuracy} \times {\early}
$$

\noindent ${\pmetric}$ evaluates the effectiveness of the predictor in predicting experts 
for prefetching by considering both ${\accuracy}$ and ${\early}$.
The ${\accuracy}$ denotes the proportion of correctly predicted experts,
while the ${\early}$ signifies the portion of predicted experts that can be prefetched in time 
before they are accessed during LLM inference.
Thus, ${\pmetric}$ measures the volume of correct experts that can be prefetched in a timely manner.

\begin{figure}[t]
\vspace{1mm}
\begin{minipage}[t]{1\linewidth}
\centering
\includegraphics[width=1\linewidth]{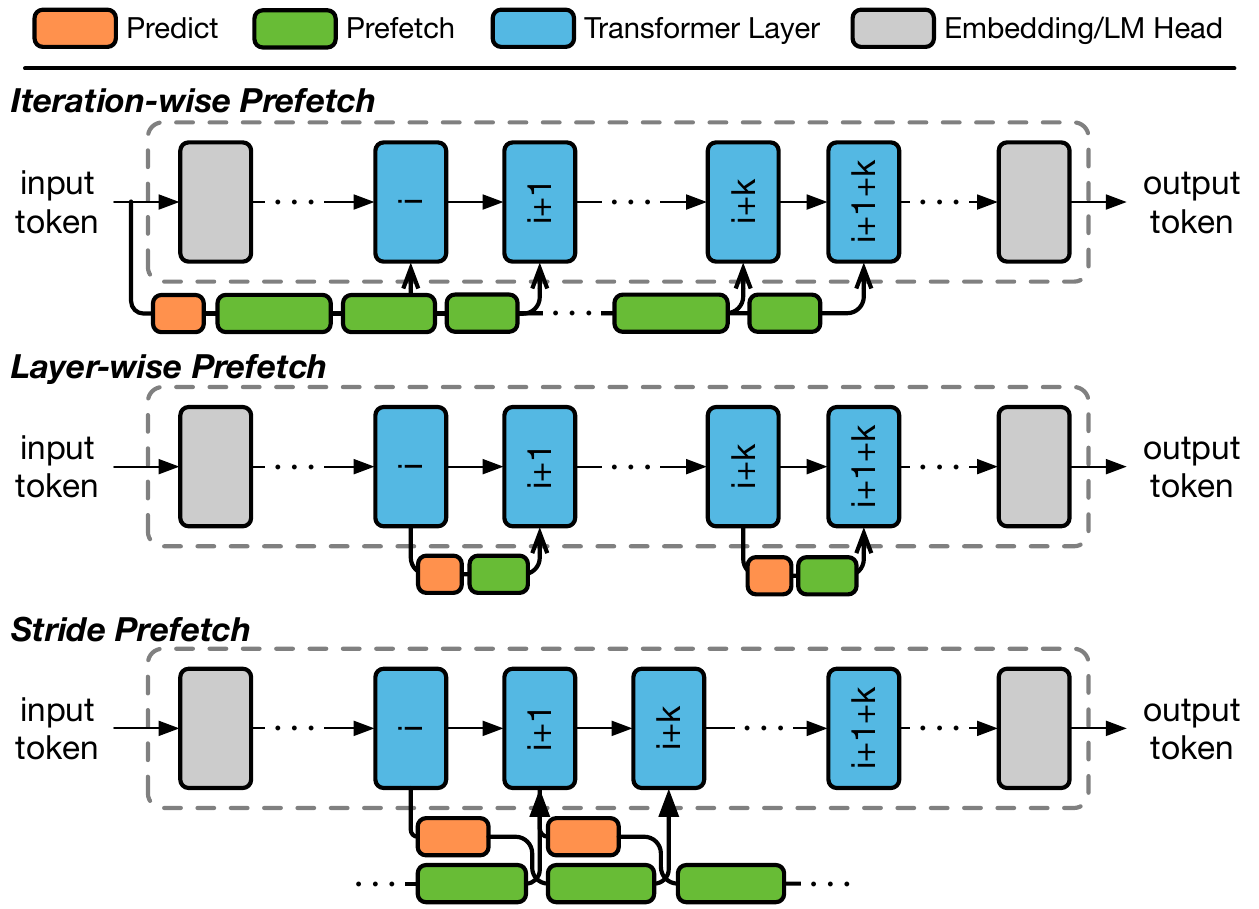}
\end{minipage} \\ [4pt]
\begin{minipage}{1\linewidth}
\caption{\small{\emph{
Candidate prefetch manners.
}}}
\label{fig:predict:methods}
\end{minipage}  \\[-15pt]
\end{figure}

\subsection{Existing Approaches}

Recent research has proposed two main methods for predicting expert usage.
Previous studies~\cite{openmoe,mixtral} introduced a \textbf{token-based} predictor that 
predicts expert usage based on input tokens, allowing for an \textbf{iteration-wise} prefetch pattern, 
as illustrated in Figure~\ref{fig:predict:methods}(a).
These studies suggest that the selection of experts in one iteration is closely related to the input token ID.
This relationship can be intuitively explained: LLMs convert the input token ID 
into an embedding vector through a fixed mapping, and the computation in each iteration 
gradually adds contextual information to these embeddings.
Consequently, the input token ID can be utilized to predict the selection of experts 
across all layers within that iteration.
Specifically, in the offline stage, a trace of input token IDs and their selected experts is collected.
Then, during online inference, the predictor determines which experts to select for one iteration
by identifying the most frequently used experts from this trace based on the input token ID.

\begin{figure}[t]
\vspace{1mm}
\begin{minipage}[t]{1\linewidth}
\begin{minipage}[t]{0.5\linewidth}
\includegraphics[width=1\linewidth]{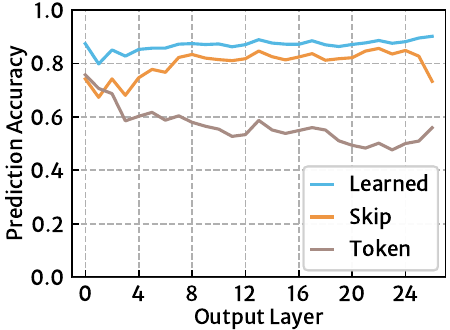}
\end{minipage}\begin{minipage}[t]{0.5\linewidth}
\includegraphics[width=1\linewidth]{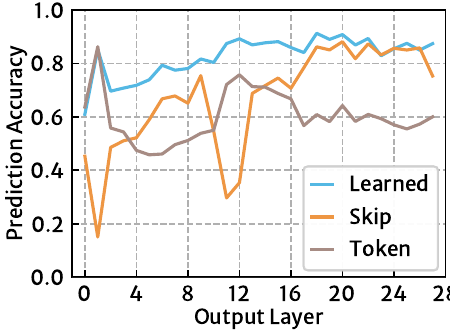}
\end{minipage}\end{minipage}\\[-1pt]
\vspace{-5pt}
\begin{minipage}[t]{1\linewidth}
\caption{\small{\emph{
The prediction accuracy of each layer in (a) DS-1 model and (b) QW-2 model.
}}}
\label{fig:predict:accuracy-all-layer}
\end{minipage}\vspace{-15pt}
\end{figure}

By predicting experts for all layers before an iteration begins, the token-based predictor
achieves optimal ${\early}$, maximizing available time for prefetching.
However, this approach suffers from low {\accuracy}.
The iteration-wise pattern conducts prediction over a long distance, leading to decreased accuracy.
Moreover, the input token ID lacks contextual information concerning the entire sequence.
As shown in Figure~\ref{fig:predict:accuracy-all-layer}, the average accuracy of the token-based predictor is only 58.3\%.
Despite delivering a high {\early} through iteration-wise prefetching,
the low {\accuracy} renders nearly half of this prefetching ineffective, resulting in a low {\pmetric}.

Another recent system~\cite{mixtral_offloading} proposed a \textbf{skip-based} predictor that facilitates a \textbf{layer-wise} prefetch manner, as illustrated in Figure~\ref{fig:predict:methods}(b).
This approach leverages the high similarity between inputs across layers in LLMs~\cite{lee_infinigen_2024,liu_deja_2023}.
It establishes a skip connection that transmits the input from $i$-th layer's MoE gate directly to the MoE gate in $i+1$-th layer, thereby predicting the experts for $i+1$-th layer at the point of $i$-th layer.
For instance, in the DS-2 model, the cosine similarity between the consecutive layers' inputs is 91.7\%.
Thus, passing the input of the $i$-th layer to the $i+1$-th layer's gate is likely to yield accurate predictions.

However, the skip-based predictor's accuracy remains limited.It depends on the similarity of inputs across different layers and the numerical stability of the gate function, which does not uniformly apply across all models.
In Figure~\ref{fig:predict:accuracy-all-layer}, the skip-based predictor achieves high accuracy with noticeable accuracy drop in the head and tail layers for the DS-1 model.
However, the QW-2 model experiences a significant accuracy decline with an average accuracy of only 66.9\%.
This discrepancy arises because the gate function in the QW-2 model is sensitive to input variations, causing shifts in priority for expert selection even with slight input changes.
Additionally, the layer-wise prefetch pattern of the skip-based predictor incurs higher prediction overhead, thus limiting prefetch efficiency.

\subsection{Learning-based Predictor}

To achieve high {\accuracy}, {\sys} introduces a learned predictor to conduct layer-wise prefetch.
The main idea is to collect the correlation between layer inputs and expert selections across layers and memorize this correlation in predictors.
The predictor then uses these correlations to make predictions.
When paired with layer-wise prefetch, the learned predictor maintains high ${\accuracy}$.

{\sys}'s learned predictor employs a small neural network (NN) to learn correlations 
between layer inputs and expert selections.
This approach, which utilizes a small NN like multi-layer perceptrons (MLPs) as the predictor, 
has been effectively applied and validated in various system research contexts~\cite{learned-index,
hao_linnos_2020,song_powerinfer_2023,liu_deja_2023}.
These NNs are capable of learning complex correlations while providing fast predictions, 
which can be more challenging 
for traditional heuristic methods like nearest neighbor search.
However, a significant drawback of NN-based methods is their lengthy training time.
Fortunately,
in the context of serving LLMs, the offline training is a one-time task for each LLM 
and is negligible compared to the extensive pre-training time required for LLMs~\cite{kaplan_scaling_2020}.

The learned predictor in {\sys} operates in two phases: offline training and online prediction.
In the offline phase, {\sys} trains a set of predictors
by performing multiple iterations of LLM inference. 
This process collects the input for each layer and the corresponding output of the gate function.
Based on these collected traces, {\sys} trains a set of predictors for each layer 
to learn and memorize the correlations between inputs and outputs.

To ensure the predictor's generalizability, {\sys} collects traces from the domain datasets 
used during either LLM training or inference.
This approach ensures that the predictor aligns with the model across various conditions.
Following standard practices, the collected traces are split into training and evaluation sets with a 9:1 ratio.
The predictor is trained solely on the training set and evaluated only on the unseen data from the evaluation set.

In the online inference, the input for each layer is collected and 
fed into the corresponding predictor(s) to make predictions.
The prediction output, similar to a gate's output, indicates the prefetch priority of experts in one layer.
Based on this output, the predictor selects the same number of experts that the model would activate for one token (e.g. 6 for the DS-1 model in Table~\ref{tbl:moe-models})
and hands over these experts to the prefetcher for prefetching.

To assess the accuracy of different predictors, we evaluated them using the evaluation set of collected traces. 
As shown in Figure~\ref{fig:predict:accuracy-all-layer}, {\sys} learned predictor maintains 
high ${\accuracy}$ across both models, achieving an average accuracy of 84.7\%.
This improved accuracy enables {\sys} to accurately prefetch experts in a timely manner, 
optimizing the use of the limited bandwidth between the CPU and GPU.

\subsection{Stride Prefetching}

To minimize the impact of the prediction on critical path latency, {\sys} executes the predictor on the CPU, allowing it to run concurrently with LLM inference.
The latency of a single predictor on the CPU is about 200 microseconds.
Compared to the millisecond-level computation time of a single layer in LLMs, this latency can be hidden since the CPU-based prediction process operates in parallel with the LLM inference on the GPU.
However, in layer-wise prefetching, the predictor's latency consumes available time for prefetching experts, resulting in a lower ${\early}$.

To enhance the ${\early}$, {\sys} introduces stride prefetching as shown in Figure~\ref{fig:predict:methods}(c).
Stride prefetching increases the prediction distance by 1, allowing prefetching to begin earlier than in layer-wise prefetching.
Moreover, stride prefetching pipelines the prediction and prefetching processes, executing them simultaneously.
In contrast to layer-wise prefetching, where prediction and prefetching are carried out sequentially,
stride prefetching ensures that all available bandwidth between the CPU and GPU is fully utilized for prefetching.
Consequently, this approach maximizes the ${\early}$ and provides a higher ${\pmetric}$.

While increasing the prediction distance may lead to a decrease in the predictor's ${\accuracy}$,
practical observations reveal that the accuracy of {\sys}'s learned predictor only declines
by 5\% during stride prefetching.
Additionally, stride prefetching offers ample design space 
for more sophisticated predictors that may require additional time to generate predictions.
 
\section{Coordination of Prefetching and Inference}
\label{sec:prefetch}

The prefetcher in {\sys} is responsible for fetching experts into the GPU cache based on prediction results.
It consists of a worker thread and a task queue.
The worker thread retrieves prefetch tasks from the queue and copies the corresponding experts into the GPU's expert cache.
The task queue maintains two priority levels: low-priority speculative prefetch tasks provided by the predictor, and high-priority precise prefetch tasks triggered by cache misses during LLM inference.
The worker thread always prioritizes executing high-priority tasks over low-priority ones.

\begin{figure}[t]
\vspace{1mm}
\begin{minipage}[t]{1\linewidth}
\centering
\includegraphics[width=1\linewidth]{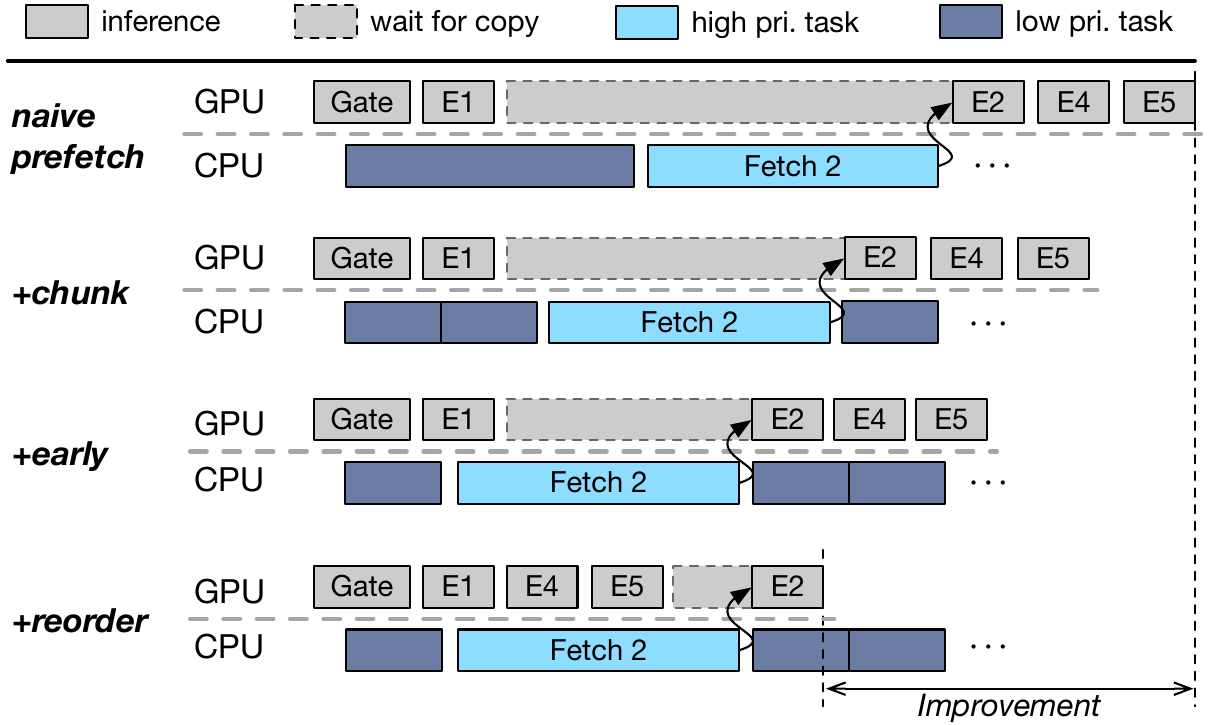}
\end{minipage} \\ [4pt]
\begin{minipage}{1\linewidth}
\caption{\small{\emph{
{\sys} coordinates prefetching with inference using a series of optimizations.
Assume experts $1,3,4,5$ are prefetched in advance, and gate produces $1,2,4,5$, where expert $2$ is not in cache.
}}}
\label{fig:prefetch:techs}
\end{minipage}  \\[-15pt]
\end{figure}

To further enhance the coordination between expert prefetching and LLM inference, 
{\sys} proposes several optimizations: chunked prefetching, early preemption, 
and reordered inference.
These optimizations aim to minimize interference and maximize the overlap 
between prefetching and inference, as illustrated in Figure~\ref{fig:prefetch:techs}.

\subsection{Chunked Prefetching}

When high-priority prefetch tasks are added to the queue, there is typically ongoing fetching of expert parameters from CPU to GPU.
This fetching may stem from an incomplete prefetch task of the current layer or from a prefetch task of subsequent layers that has already begun.
Due to the limitations of CUDA's asynchronous copy mechanism, an ongoing copy operation cannot be preempted midway.
As a result, high-priority prefetch tasks must wait for the current copy operation to complete before they can start,
introducing unnecessary latency into the critical path.

To address this issue, {\sys} introduces \emph{chunk-based prefetching}.
The key idea is to split the parameters of each expert into multiple chunks.
When the prefetcher identifies the predicted experts (from predictor), 
it divides their parameters into several chunks and adds them to the prefetch queue 
as low-priority tasks.
Each task corresponds to one chunk of an expert's parameters, rather than the entire expert.
This allows the worker thread to schedule low-priority tasks at a smaller granularity.
When a high-priority prefetch task arises, the worker thread can quickly switch to it, 
encountering a maximum delay of just one chunk.

Figure~\ref{fig:prefetch:techs} illustrates an example of chunked prefetching.
The cache miss for expert $2$ is triggered after the execution of expert $1$.
Since the prefetcher is already working on a low-priority task, it must wait until this task completes before handling the high-priority task of expert $2$.
With chunked prefetching, the low-priority task is divided into three chunks.
The cache miss is triggered while the prefetcher is working on the second chunk, allowing the high-priority task of expert $2$ to start immediately after the second chunk is completed.
In practice, we found that experts in MoE models share the same structure, consisting of three linear layers.
Thus, {\sys} naturally splits each expert into three chunks, corresponding to these three linear layers.
By implementing chunked prefetching, {\sys} reduces the delay in starting high-priority prefetch tasks, thereby improving critical path latency.

\subsection{Early Preemption}
Although {\sys}'s predictor aims to maximize prediction accuracy, mispredictions are still unavoidable.
This can result in necessary experts not being present in the GPU cache, triggering on-demand copying of missing experts during the critical path.
Traditionally, these misses are detected and addressed only when the corresponding expert is accessed during inference,
causing the inference process to be blocked while waiting for the missing expert parameters to be copied from CPU memory to GPU.
This leads to under-utilization of the GPU and introduces high fetch latency in the critical path of inference execution.

To tackle this issue, {\sys} proposes \emph{early preemption}. We observed that, in MoE models, the experts needed for the current layer are determined 
all at once when the gate function completes.
Instead of causing a cache miss each time an individual expert is accessed, 
the system can preempt the prefetch queue in advance when it knows 
which experts will be activated after the gate function.
This allows the prefetching of any missing experts to begin much earlier, 
overlapping with the computation of the current layer.
For example, as shown in Figure~\ref{fig:prefetch:techs}, early preemption triggers the cache miss for expert $2$ immediately 
after the gate function completes, rather than waiting until the completion of expert $1$.
As a result, the high-priority task for expert $2$ is scheduled by 
the prefetcher before the second chunk of the low-priority task is processed.

In practice, {\sys} implements early preemption by inserting a hook 
at the end of the gate function to obtain the list of required experts in advance.
These experts are then prioritized as high-priority tasks and added to the prefetch queue, 
ensuring that the prefetch thread prioritizes these tasks.
During this process, there may still be some low-priority speculative prefetch tasks 
for the same layer that have not yet completed.
However, since the system has a precise list of the required experts, 
these low-priority tasks can be discarded.
The prefetch thread simply clears any remaining low-priority speculative prefetch tasks 
for that layer, effectively achieving preemption.

During inference, when encountering an expert that is not in the cache, {\sys} no longer triggers a cache miss.
Instead, it waits for the corresponding prefetch task to complete.
As a result, all passive cache misses are transformed into proactive precise prefetching.
This approach allows for earlier initiation of accurate prefetching, which increases the overlap between prefetching and computation, ultimately reducing latency on the critical path.

\begin{algorithm}[t]
\caption{Prefetch Worker Thread}
\label{alg:prefetch-worker}
\begin{algorithmic}[1]
\While{True}
    \State task $\gets$ queue.pop()
    \If{task.chunk = 0}
        \State evicted\_expert $\gets$ cache.replace\_with(task.expert)
        \State evicted\_expert.ready\_chunk $\gets$ 0
    \EndIf
    \State cache\_ptr $\gets$ cache.get(task.expert)
    \State offset $\gets$ task.chunk $\times$ chunk\_size
    \State copy(cache\_ptr + offset, task.host\_ptr + offset, chunk\_size)
    \State task.expert.ready\_chunk $\gets$ task.chunk + 1
\EndWhile
\vspace{0.5mm}
\end{algorithmic}
\end{algorithm}

\subsection{Reordered Inference}
\label{sec:prefetch:reorder}
In the inference process of LLMs, existing frameworks typically execute computations for different experts in the order of their IDs.
This order often fails to fully utilize the cache status of experts, leading to unnecessary blocking and potential cache thrashing.
Consider the example in Figure~\ref{fig:prefetch:techs} where experts $1$, $4$, and $5$ are cached, 
and expert $2$ is missing.
Since the computations are executed based on the order of expert ID, experts $4$ and $5$ must wait for 
the prefetch of expert $2$ to complete before they can start.
Consequently, the GPU remains underutilized while waiting for the prefetch of expert $2$, 
even though experts $4$ and $5$ are already prefetched.
More critically, the prefetching of the missing expert might evict other soon-to-be-accessed experts, causing cache thrashing.
This issue is particularly severe when dealing with a large number of experts sequentially, such as during the prefill stage of inference.

To address this issue, {\sys} proposes \emph{reordered inference}, 
which alters the computation order of experts in a cache-aware manner.
We observe that in MoE models, the computation order of experts is interchangeable.
There is no dependency between the computations of different experts because their outputs are simply summed together.
This property allows for adjusting the computation order based on the cache and prefetch status, making the inference process more cache-friendly.

Specifically, once the gate function completes, {\sys} adjusts the computation order accordingly.
Experts already in the cache are prioritized first, followed by the experts currently being prefetched (if any), while experts whose prefetch has not yet begun are positioned last.
Consider the example in Figure~\ref{fig:prefetch:techs}.
When the gate produces experts $1$, $2$, $4$, and $5$, where expert $2$ is missing, 
{\sys} changes the computation order to $1$, $4$, $5$, and then $2$.
Therefore, the prefetching of expert $2$ can be conducted in parallel with the computations of experts $4$ and $5$, 
further reducing the impact of prefetching on the critical path.

In practice, the reordering process occurs simultaneously with early preemption.
After obtaining the list of experts to be accessed, {\sys} first reorders them as described above.
Experts whose prefetching is not yet complete are managed through early preemption and added to the prefetch queue as high-priority tasks.
The entire reordered sequence of experts is then returned to the inference framework for execution.
This approach ensures that for experts with incomplete prefetches, both the prefetch threads and inference threads process them in the same order, effectively establishing a pipeline between computation and prefetching.

\begin{algorithm}[t]
\caption{Prefetcher Interface}
\label{alg:prefetch-interface}
\begin{algorithmic}[1]
\Function{PushPredictedExperts}{layer, experts}
    \For{e \textbf{in} experts}
        \If{e.ready\_chunk > 0}
            \State cache.hit(e)
        \EndIf
        \For{chunk $\gets$ e.ready\_chunk \textbf{to} num\_chunks-1}
            \State queue.push(Task(layer, e, chunk, LOW))
        \EndFor
    \EndFor
\EndFunction

\Function{PushPreciseExperts}{layer, experts}
    \State queue.remove\_low\_pri\_task\_with\_layer(layer)\State experts $\gets$ desc\_sort\_by\_ready\_chunk(experts)\For{e \textbf{in} experts}
        \If{e.ready\_chunk > 0}
            \State cache.hit(e)
        \EndIf
        \For{chunk $\gets$ e.ready\_chunk \textbf{to} num\_chunks-1}
            \State queue.push(Task(layer, e, chunk, HIGH))
        \EndFor
    \EndFor
    \State \Return experts
\EndFunction
\vspace{0.5mm}
\end{algorithmic}
\end{algorithm}

\subsection{Prefetcher Workflow}
\label{sec:prefetch:workflow}

The prefetcher's workflow is summarized in Algorithms~\ref{alg:prefetch-worker} and~\ref{alg:prefetch-interface}.
Algorithm~\ref{alg:prefetch-worker} outlines the prefetcher's worker thread, which continuously polls tasks from the queue and transfers expert parameters from host memory to GPU memory.
Each task corresponds to a chunk of an expert's parameters, thereby implementing {chunked prefetching}.

The Predictor and LLM framework interact with the prefetcher through the APIs 
outlined in Algorithm~\ref{alg:prefetch-interface}.
The Predictor enqueues predicted experts as low-priority tasks using the {\tt PushPredictedExperts} function,
while the LLM framework enqueues the actually required (precise) experts as high-priority tasks 
with the {\tt PushPreciseExperts} function after completing the gate function.

When enqueuing high-priority tasks (precise experts), the system first clears 
any existing low-priority tasks from the queue (Line 12) to enable early preemption.
The remaining precise experts are then reordered based on their current fetch status (Line 13).
Subsequently, the inference framework executes the experts according to this new ordering (Line 22), 
thereby implementing reordered inference.
 
\section{Implementation}
\label{sec:impl}

{\sys} is implemented as an extension to LLM frameworks, comprising 6,600 lines of C++ code.

\subsection{Cache Implementation}
For simplicity, the cache component of {\sys} is implemented as a standard per-layer LRU cache.
Both prefetching and inference trigger a cache access.
When adding prefetch tasks, {\sys} leverages LRU by accessing experts that are already cached, thereby delaying their eviction.
To reduce memory fragmentation, {\sys} pre-allocates the expert cache as a contiguous memory region.

\subsection{System Integration}
We have integrated {\sys} into two popular LLM frameworks: transformers and llama.cpp.
To achieve this integration, we added hooks to capture input logits from the MoE layers and to reorder experts.
We also implemented a dependency mechanism to ensure efficient prefetching and computation.
Furthermore, {\sys} takes over the memory management for expert parameters.
We did not integrate {\sys} with frameworks like vLLM and TGI due to their inadequate support for quantized MoE at the time of submission.
Moreover, these frameworks primarily focus on batched inference for data centers and fuse the execution of multiple experts to enhance GPU utilization.
However, this optimization assumes that all experts are ready before computation can begin, which is atypical on memory-constrained GPU platforms that {\sys} targets.
It necessitates additional GPU memory for activated experts and hampers the overlap between expert loading and execution.

\subsection{Training of Predictor}
Each layer's predictor in {\sys} is implemented as a two-layer multi-layer perceptron (MLP) 
with approximately 2 million parameters.
The training of the learned predictor and the data collection for training 
takes less than 1--2 hours on a single GPU.
This is a one-time offline task that can be parallelized across multiple GPUs.
Given the extensive pre-training times of large language models (LLMs), 
we consider this time commitment acceptable.

\section{Evaluation}
\label{sec:eval}

\subsection{Experimental Setup}
\label{sec:setup}

\nospacestitle{Hardware}. The evaluation is conducted on a PC equipped with an NVIDIA RTX 4090 GPU (24\,GB GDDR6X).
The PC also features an Intel i9-14900K CPU and 128\,GB of host DRAM.
The GPU is connected to the CPU via PCIe 4.0, providing a unidirectional bandwidth of 32\,GB/s.
To simulate GPUs with varying memory capacities, we include an evaluation 
in $\S$\ref{sec:eval:sens-cache-rate} that adjusts the cache ratio to control memory occupancy.

\stitle{Workload}.
We evaluated a broad range of MoE-based LLMs, as listed in Table~\ref{tbl:moe-models}.
By default, we evaluate DS-1, DS-2, and QW-1 using FP16 precision, while QW-2 and Mixt are evaluated using INT4 precision.
To further study the impact of model size, we also include an evaluation 
in $\S$\ref{sec:eval:sens-model-size} that varies the parameter size of the same model from FP16 to INT4.
The evaluation utilizes the shareGPT dataset \cite{sharegpt}, which consists of user interactions with ChatGPT and serves as a representative example of real LLM services.
We also conducted evaluations using the Alpaca dataset \cite{alpaca} and observed similar performance trends;
results for this dataset are omitted due to space limitations.
By default, we set the batch size to 1 to reflect edge deployment scenarios,
and we include an evaluation in $\S$\ref{sec:eval:sens-bs} that varies the batch size from 1 to 4.

\stitle{Baselines}.
Our evaluation relies on two well-known codebases: Hugging Face transformers~\cite{transformers,accelerate} and llama.cpp~\cite{llamacpp}.
Transformers supports a wide range of models and is easy to deploy, but it lacks optimal inference performance.
We enhanced the efficiency of the MoE block by reducing CPU-GPU synchronization.
Llama.cpp, which is written in C++, delivers state-of-the-art inference performance by eliminating overhead from the Python interpreter.

Both systems offer offloading baselines: transformers offloads only the parameters to the CPU (referred to as \textbf{TO}), while llama.cpp offloads both parameters and computations (referred to as \textbf{LO}).
We improved TO by incorporating pinned memory and asynchronous copies.
Additionally, we integrated {\sys} into both codebases and introduced three baselines: Unified Memory (\textbf{UM}), \textbf{static} cache, and \textbf{LRU} cache.
These baselines, along with {\sys}, manage only expert parameters, while non-expert parameters consistently reside on the GPU.
The UM baseline is optimized using \texttt{cudaMemAdvise} to enable instantaneous page invalidation without incurring the cost of swapping pages back to CPU memory.
The static baseline caches a fixed set of experts and utilizes two additional expert buffers to load any missing experts.

\begin{figure}[t]
\begin{minipage}[t]{1\linewidth}
\centering
\includegraphics[width=1\linewidth]{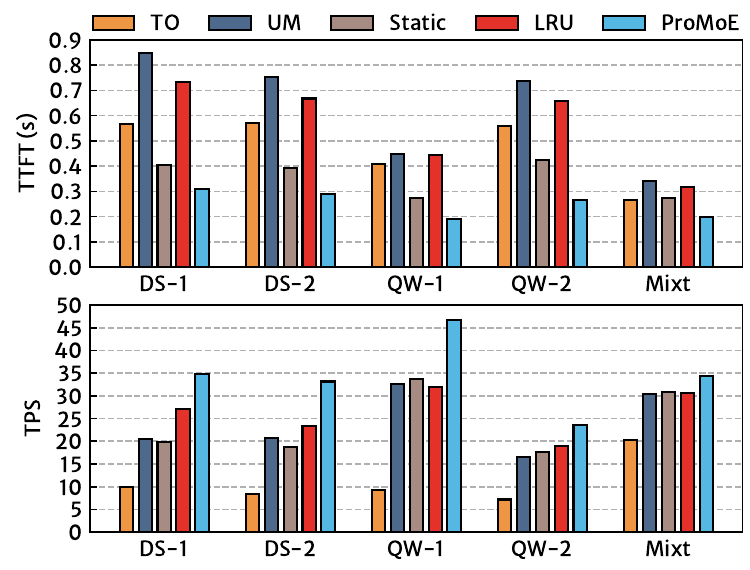}
\end{minipage} \begin{minipage}{1\linewidth}
\caption{\small{\emph{
The overall performance of (a) prefill and (b) decode stage in transformers codebase.}}}
\label{fig:eval:overall-transformers}
\end{minipage}  \\[-18pt]
\end{figure}

\stitle{Metrics}
We evaluate the performance of {\sys} and its baselines in the prefill and decode stages separately.
The prefill stage performance is assessed by TTFT (Time To First Token), 
which reflects the latency in processing the user's prompt.
The decode stage performance is measured using TPS (Tokens Per Second) and TPOT (Time Per Output Token), 
indicating the throughput and latency of the decoding process.
We primarily report TPS as it is more intuitive and switch to TPOT for detailed breakdown analyses.
The total latency for a single request can be expressed 
as $Latency_{total} = \text{TTFT} + N \times \text{TPOT}$ (where $N$ is output length).
We measure the prefill and decode stages separately for two main reasons: 
(1) the significant variance in output lengths (ranging from tens to thousands of tokens) 
renders aggregated metrics unreliable for system comparisons, 
and (2) the prefill and decode stages exhibit distinct computational patterns
(e.g. more experts are activated in the prefill stage).

\subsection{Overall Performance}
Figure~\ref{fig:eval:overall-transformers} shows the overall performance of the prefill 
and decode stages within the transformers codebase.
In the prefill stage, {\sys} outperforms static and LRU baselines by an average of 1.42$\times$ (up to 1.61$\times$) and 2.21$\times$ (up to 2.48$\times$), respectively.
The improvement of {\sys} primarily stem from its prefetching technique, 
which maximizes the overlap between loading parameters and computation.
When comparing {\sys} with LRU, the greater improvement is attributed to the cache thrashing caused by LRU
(see $\S$\ref{sec:prefetch:reorder}).
In the prefill stage, nearly all experts are accessed since each token usually requires a different set of experts.
As experts are accessed according to their IDs, LRU evicts a cached expert with a higher ID when it accesses a missing expert with a smaller ID first.
The static cache avoids thrashing by fixing its cache set, while {\sys} intelligently reorders experts to minimize thrashing and reduce the blocking time caused by missing experts on the critical path.

In the decode stage, {\sys} outperforms the static and LRU baselines by an average of 1.47$\times$ (up to 1.77$\times$) and 1.31$\times$ (up to 1.46$\times$), respectively.
LRU outperforms the static cache during the decode stage because cache thrashing occurs less frequently, and there is some reuse of experts across iterations.
{\sys} excels over these baselines by keeping most copies of missing experts off the critical path through effective prefetching.

The TO (resp. UM) baseline consistently perform worse than the static (resp. LRU) baseline.
Comparing to these baselines, {\sys} achieves a average speedup of 2.15$\times$ (up to 2.78$\times$) in the prefill stage and 2.47$\times$ (up to 5.02$\times$) in the decode stage.
This performance gap arises because the static and LRU baselines can be seen as improved implementations of static and dynamic cache, respectively.
In static cache, the TO baseline offloads non-expert parameters to the CPU, while the static baseline only offloads parameters of the experts.
In dynamic cache, the UM baseline fetches parameters at the page level, which increases the volume of transferred data compared to the LRU baseline. 
Therefore, in subsequent experiments, we mainly focus on comparing the static, LRU, and {\sys}.

Figure~\ref{fig:eval:overall-llama.cpp} shows the overall performance in the llama.cpp codebase.
{\sys} surpasses the static and LRU baselines by an average of 1.36$\times$ (up to 1.75$\times$) and 2.12$\times$ (up to 2.22$\times$) in the prefill stage, and by 1.49$\times$ (up to 1.79$\times$) and 1.09$\times$ (up to 1.17$\times$) in the decode stage, respectively.
The improvement in the llama.cpp codebase follows the same trend observed in transformers.
However, it is less pronounced due to the removal of the Python interpreter overhead during inference, which provides fewer opportunities for {\sys} to prefetch experts.

\diffatomic{
\begin{figure}[t]
\begin{minipage}[t]{1\linewidth}
\centering
\includegraphics[width=1\linewidth]{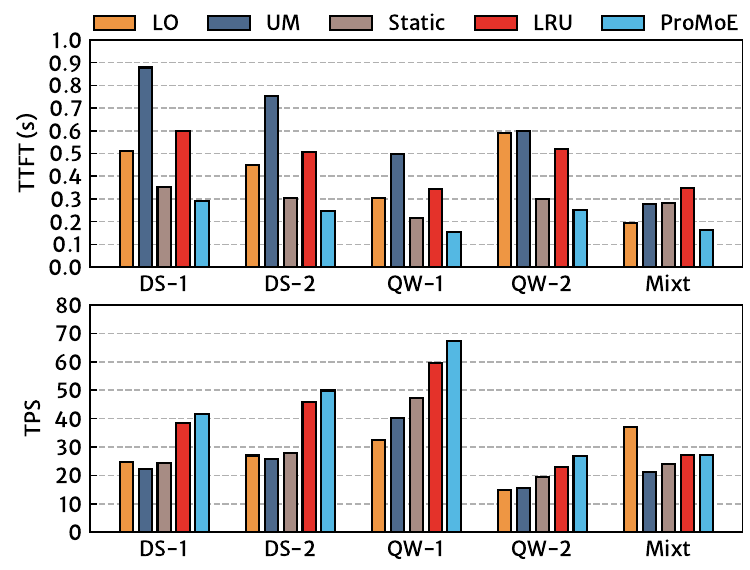}
\end{minipage} \begin{minipage}{1\linewidth}
\caption{\small{\emph{
The overall performance of (a) prefill and (b) decode stage in llama.cpp codebase.}}}
\label{fig:eval:overall-llama.cpp}
\end{minipage}  \\[-18pt]
\end{figure}
}

As expected, the UM baseline consistently performs worse than the LRU baseline.
The LO baseline in llama.cpp offloads both parameters and computations to the CPU, resulting in slower performance than that of the static baseline.
Compared to these baselines, {\sys} achieves an average speedup of 2.25$\times$ (up to 3.21$\times$) in the prefill stage and 1.66$\times$ (up to 2.07$\times$) in the decode stage.
However, when evaluating the Mixt model, the LO baseline is significantly faster and even surpasses {\sys} in the decode stage.
This is because the Mixt model activates a larger ratio of experts (25\%) for each token, increasing the cost of fetching parameters to the GPU compared with directly computing them on the CPU.
We believe this does not undermine the significance of our work, as most recently released MoE-based LLMs typically activate a smaller ratio of experts (averaging 10\%), and the TO baseline continues to show inferior performance across most cases.

\begin{figure}[t]
\begin{minipage}[t]{1\linewidth}
\centering
\includegraphics[width=1\linewidth]{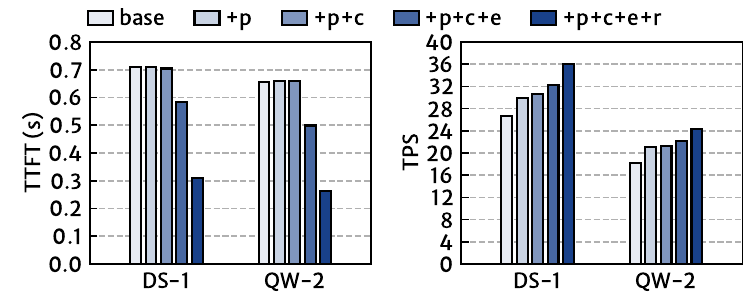}
\end{minipage} \begin{minipage}{1\linewidth}
\caption{\small{\emph{
The ablation study of (a) prefill and (b) decode stage in transformers codebase with different optimizations in {\sys} enabled.
Base represents the LRU baseline, with \underline{p}refetch, \underline{c}hunked-prefetch, \underline{e}arly-preemption and \underline{r}eordered-inference applied.
}}}
\label{fig:eval:ablation-transformers}
\end{minipage}  \\[-8pt]
\end{figure}
\begin{figure}[t]

\begin{minipage}[t]{1\linewidth}
\centering
\includegraphics[width=1\linewidth]{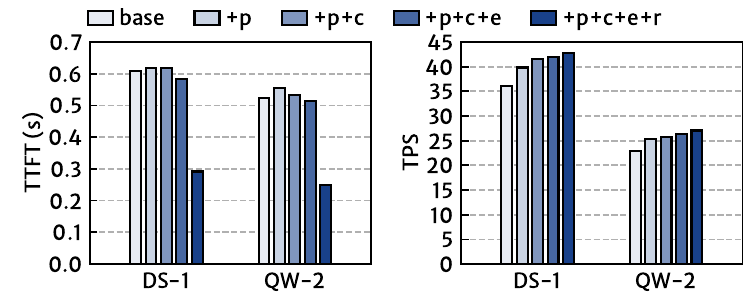}
\end{minipage} \begin{minipage}{1\linewidth}
\caption{\small{\emph{
The ablation study in llama.cpp codebase, following the same setup and naming convention as Figure~\ref{fig:eval:ablation-transformers}.}}}
\label{fig:eval:ablation-llama.cpp}
\end{minipage}  \\[-10pt]
\end{figure}

\subsection{Ablation Study}
Figure~\ref{fig:eval:ablation-transformers} shows the performance of transformers 
with different optimizations enabled in {\sys}, starting from the LRU baseline.
During the prefill stage, enabling prefetching shows minimal improvement and may even degrade performance.
This is because nearly all experts are accessed, and prefetching alone merely replaces the cache set.
Additionally, naive prefetching delays the handling of missing experts.
The techniques of early preemption and reordered inference provide significant improvements,
yielding speedups of 1.27$\times$ and 2.39$\times$ compared to the baseline, respectively.
In the decode stage, these techniques gradually enhance performance, 
resulting in a 1.35$\times$ increase over the baseline.
Figure~\ref{fig:eval:ablation-llama.cpp} presents an ablation study for the llama.cpp. The trends is similar to those observed in the transformers, except that in the prefill stage, most of the improvement is attributed to the reordered inference.

\subsection{Impact of Cache Rate}
\label{sec:eval:sens-cache-rate}
\begin{figure}[t]
\begin{minipage}[t]{1\linewidth}
\centering
\includegraphics[width=1\linewidth]{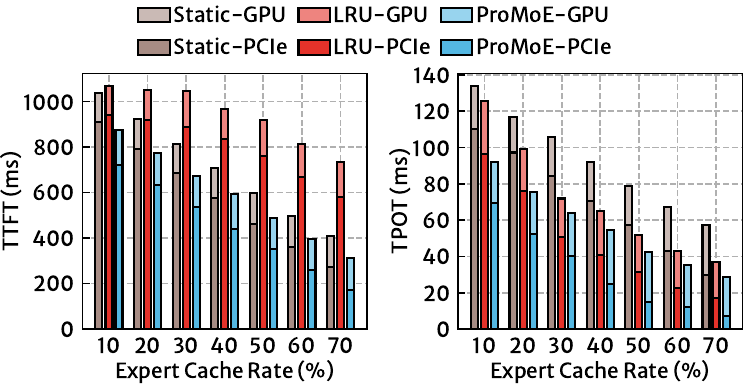}
\end{minipage} \begin{minipage}{1\linewidth}
\caption{\small{\emph{
The (a) TTFT and (b) TPOT of systems in transformers codebase with DS-1 model using different cache rates.}}}
\label{fig:eval:sens-cache-rate-trans-ds1}
\end{minipage}  \\[-8pt]
\end{figure}

\begin{figure}[t]
\begin{minipage}[t]{1\linewidth}
\centering
\includegraphics[width=1\linewidth]{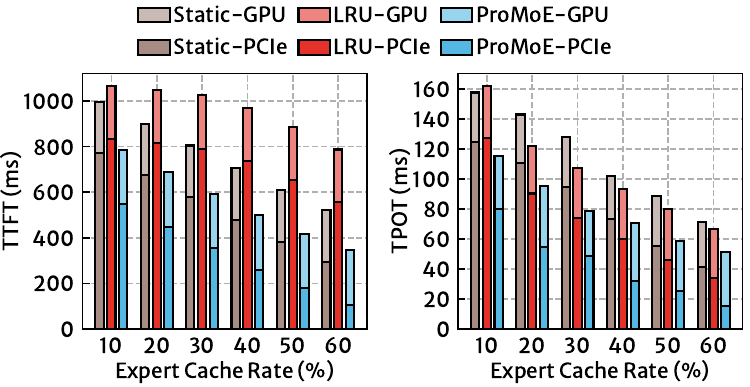}
\end{minipage} \begin{minipage}{1\linewidth}
\caption{\small{\emph{
The (a) TTFT and (b) TPOT of systems in transformers codebase with QW-2 model using different cache rates.}}}
\label{fig:eval:sens-cache-rate-trans-qw2}
\end{minipage}  \\[-8pt]
\end{figure}

To examine the impact of GPU memory capacity on {\sys}'s performance, we varied the cache rate to control the memory occupied by the expert cache.
Figure~\ref{fig:eval:sens-cache-rate-trans-ds1} and~\ref{fig:eval:sens-cache-rate-trans-qw2} show the performance of prefill and decode stages of systems in the transformers codebase, with DS-1 and QW-2 models using different cache rates.
During the prefill stage, LRU performs the worst due to cache thrashing, while {\sys} outperforms LRU on the DS-1 and QW-2 models by 1.72$\times$ (up to 2.36$\times$) and 1.82$\times$ (up to 2.28$\times$) on average, respectively.
Compared to static caching, {\sys} achieves speedups of 1.22$\times$ and 1.39$\times$ on average in the prefill stages of these two models, respectively.
The enhancement in the QW-2 model is more pronounced due to its increased computation during inference, allowing more opportunities for {\sys} to prefetch experts.
Figure~\ref{fig:eval:sens-cache-rate-trans-qw2} also shows the breakdown of time spent loading parameters on the critical path.
{\sys} reduces the loading time on the critical path from 69.68\% to 30.96\% in the QW-2 model as the cache rate increases, whereas the static cache still suffers from a reduction of only 77.44\% to 56.04\%.
In the decode stage, {\sys} outperforms both static and LRU baselines by 1.60$\times$ and 1.29$\times$ on average, respectively.
{\sys} decreases the loading time on the critical path to 25.61\% and 29.20\% for the DS-1 and QW-2 models, while LRU (the fast baseline) continues to endure loading times on the critical path of 45.52\% and 50.89\%, respectively.

\begin{figure}[t]
\begin{minipage}[t]{1\linewidth}
\centering
\includegraphics[width=1\linewidth]{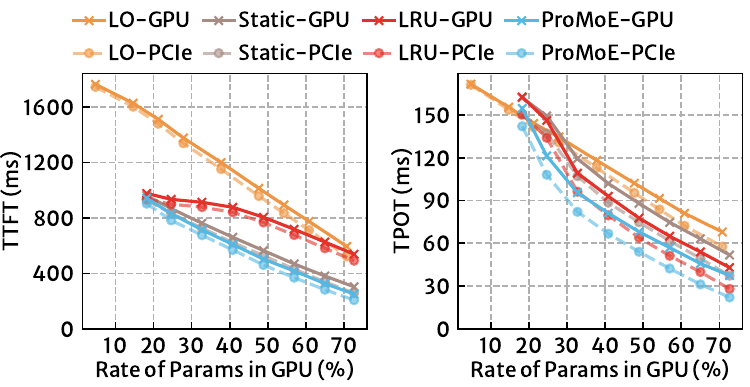}
\end{minipage} \begin{minipage}{1\linewidth}
\caption{\small{\emph{
The (a) TTFT and (b) TPOT of systems in llama.cpp codebase with QW-2 model using different cache rates.}}}
\label{fig:eval:sens-cache-rate-llama-qw2}
\end{minipage}  \\[-10pt]
\end{figure}

We conducted similar experiments on the llama.cpp codebase, using layer-offloading (LO) included as a baseline.
The results are shown in Figure~\ref{fig:eval:sens-cache-rate-llama-qw2}.
In this case, the speedup of {\sys} over the fast baseline is reduced due to the faster inference speed of the llama.cpp codebase.
{\sys} achieves performance improvements of 1.53$\times$ (resp. 1.14$\times$) 
and 1.10$\times$ (resp. 1.27$\times$) over LRU and static baselines on average 
during the prefill (resp. decode) stage, respectively.
Notably, in the decode stage with a low cache rate, LO outperforms the other systems.
Under low cache rates, the cache-based systems must fetch a significant number of experts through PCIe, 
while the limited computation makes offloading to the CPU more advantageous.
As the cache rate increases, however, the cache-based systems quickly surpass LO.

\subsection{Impact of Batch Size}
\label{sec:eval:sens-bs}
\begin{figure}[t]
\begin{minipage}[t]{1\linewidth}
\centering
\includegraphics[width=1\linewidth]{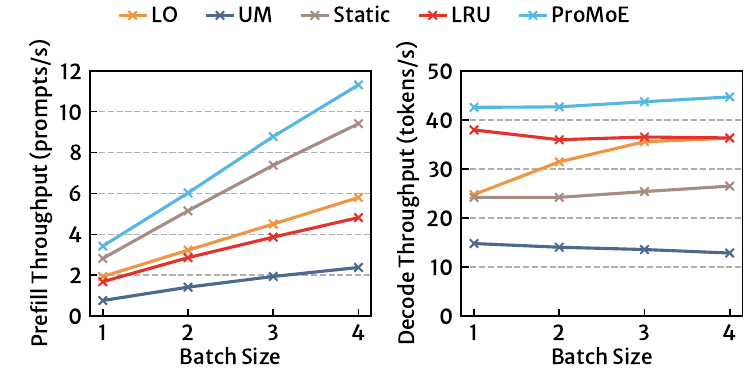}
\end{minipage} \begin{minipage}{1\linewidth}
\caption{\small{\emph{
The (a) prefill and (b) decode throughput of systems in llama.cpp codebase with DS-1 model as the batch size changes.}}}
\label{fig:eval:sens-bs-llama-thpt-ds1}
\end{minipage}  \\[-8pt]
\end{figure}

\begin{figure}[t]
\begin{minipage}[t]{1\linewidth}
\centering
\includegraphics[width=1\linewidth]{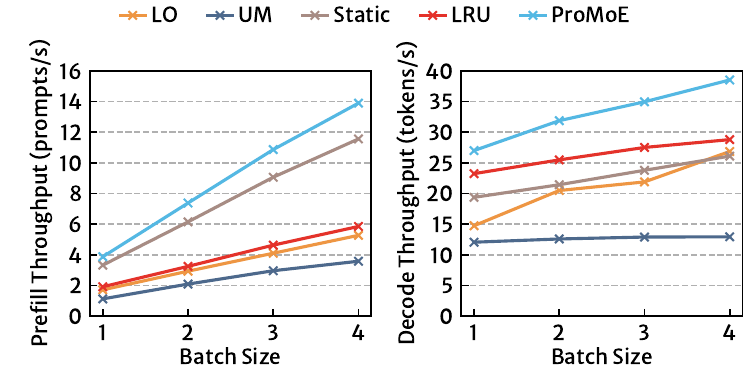}
\end{minipage} \begin{minipage}{1\linewidth}
\caption{\small{\emph{
The (a) prefill and (b) decode throughput of systems in llama.cpp codebase with QW-2 model as the batch size changes.}}}
\label{fig:eval:sens-bs-llama-thpt-qw2}
\end{minipage}  \\[-8pt]
\end{figure}

\begin{figure}[t]
\begin{minipage}[t]{1\linewidth}
\centering
\includegraphics[width=1\linewidth]{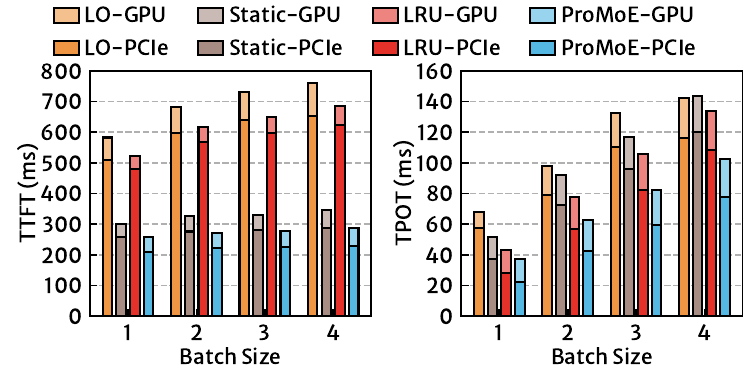}
\end{minipage} \begin{minipage}{1\linewidth}
\caption{\small{\emph{
The (a) TTFT and (b) TPOT of systems in llama.cpp codebase with QW-2 model as the batch size changes.}}}
\label{fig:eval:sens-bs-llama-time-qw2}
\end{minipage}  \\[-10pt]
\end{figure}

We also evaluated the impact of batch size on the performance of {\sys}.
Figure~\ref{fig:eval:sens-bs-llama-thpt-ds1} and~\ref{fig:eval:sens-bs-llama-thpt-qw2} show the throughput of systems in the llama.cpp codebase with DS-1 and QW-2 models as the batch size varies.
During the prefill stage, throughput increases linearly with the batch size.
This linear growth occurs because the time is primarily dominated by loading all experts, and the increased computation associated with a larger batch size is almost ``free''.
This is supported by Figure~\ref{fig:eval:sens-bs-llama-time-qw2}(a), which shows the time breakdown of the prefill stage for the QW-2 model.
As the batch size increases, the latency for one iteration in the prefill stage remains relatively stable.
On average, {\sys} outperforms LRU and static baselines by 2.19$\times$ and 1.19$\times$, respectively, in the prefill stage.

In the decode stage, the number of experts activated grows almost linearly with the batch size.
This rapid increase in latency per iteration during the decode stage limits the improvement of throughput as the batch size increases.
In this context, {\sys} outperforms both the LRU and static baselines by averages of 1.22$\times$ and 1.59$\times$, respectively.
The improvement of {\sys} over LRU grows progressively with increasing batch sizes.
For instance, in the QW-2 model, the speedup of {\sys} over LRU is 1.16$\times$ when the batch size is 1 and increases to 1.34$\times$ when the batch size reaches 4.
This improvement is attributed to cache thrashing that occurs as the batch size grows.

\begin{figure}[t]
\begin{minipage}[t]{1\linewidth}
\centering
\includegraphics[width=1\linewidth]{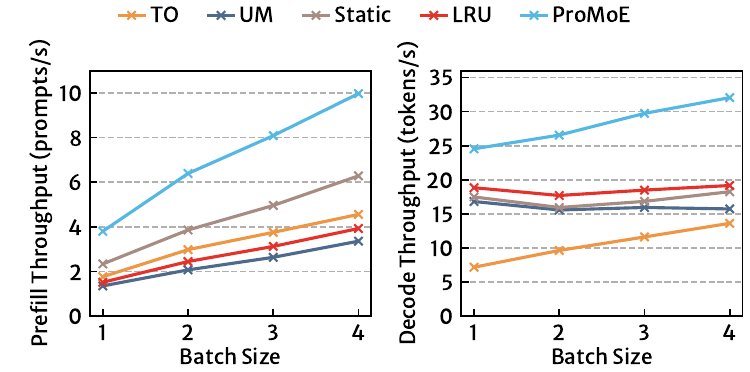}
\end{minipage} \begin{minipage}{1\linewidth}
\caption{\small{\emph{
The (a) prefill and (b) decode throughput of systems in transformers codebase with QW-2 model as the batch size changes.}}}
\label{fig:eval:sens-bs-transformers-thpt-qw2}
\end{minipage}  \\[-8pt]
\end{figure}

\begin{figure}[t]
\begin{minipage}[t]{1\linewidth}
\centering
\includegraphics[width=1\linewidth]{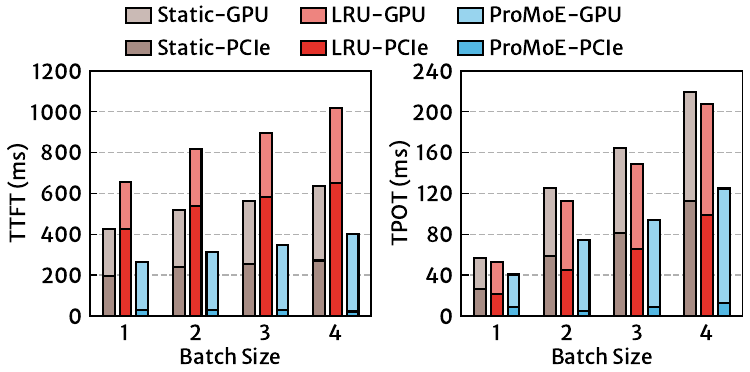}
\end{minipage} \begin{minipage}{1\linewidth}
\caption{\small{\emph{
The (a) TTFT and (b) TPOT of systems in transformers codebase with QW-2 model as the batch size changes.}}}
\label{fig:eval:sens-bs-transformers-time-qw2}
\end{minipage}  \\[-8pt]
\end{figure}

We further illustrate the impact of batch size in the transformers codebase in Figure~\ref{fig:eval:sens-bs-transformers-thpt-qw2} and~\ref{fig:eval:sens-bs-transformers-time-qw2}.
Here, {\sys} outperforms LRU and static baselines by averages of 2.47$\times$ (1.48$\times$) and 1.54$\times$ (1.87$\times$) in the prefill (decode) stage, respectively.
The higher speedup is a result of longer computation times in the transformers codebase, which provides {\sys} with more opportunities to perform additional prefetches.

\begin{figure}[t]
\begin{minipage}[t]{1\linewidth}
\centering
\includegraphics[width=1\linewidth]{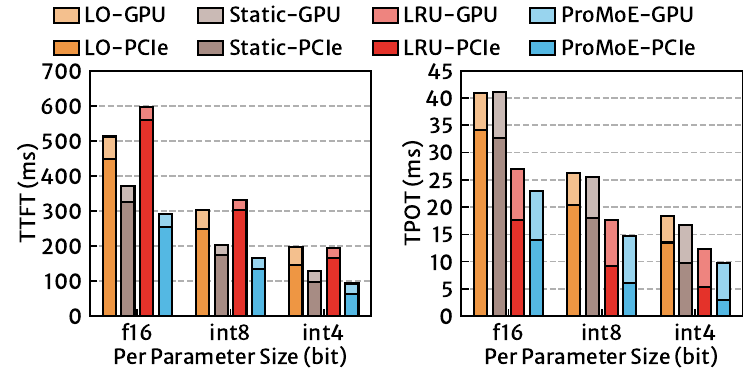}
\end{minipage} \begin{minipage}{1\linewidth}
\caption{\small{\emph{
The (a) TTFT and (b) TPOT of systems in llama.cpp codebase with DS-1 model using different bits per weight.}}}
\label{fig:eval:sens-model-size-llama-ds1}
\end{minipage}  \\[-10pt]
\end{figure}

\subsection{Impact of Model Size}
\label{sec:eval:sens-model-size}
To understand how model size affects {\sys}'s performance, we varied the number of bits per weight (BPW) from 16 to 4 for the same model.
The variance in BPW impacts the model's total memory footprint while keeping the amount of computation, measured in FLOPs (floating-point operations), constant.
This variability can alter the relative speed of parameter loading and computation.

\begin{figure}[t]
\vspace{1mm}
\begin{minipage}[t]{1\linewidth}
\centering
\includegraphics[width=1\linewidth]{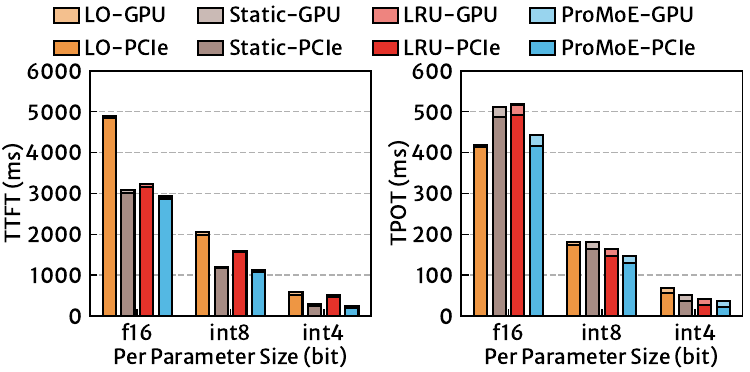}
\end{minipage} \begin{minipage}{1\linewidth}
\caption{\small{\emph{
The (a) TTFT and (b) TPOT of systems in llama.cpp codebase with QW-2 model using different bits per weight.}}}
\label{fig:eval:sens-model-size-llama-qw2}
\end{minipage}  \\[-8pt]
\end{figure}

Figure~\ref{fig:eval:sens-model-size-llama-ds1} and~\ref{fig:eval:sens-model-size-llama-qw2} present the results for the DS-1 and QW-2 models within the llama.cpp codebase.
In the DS-1 model, we maintained a fixed cache rate to keep a consistent ratio of parameters stored in the GPU.
The GPU memory occupancy decreases with lower BPW values.
As shown in Figure~\ref{fig:eval:sens-model-size-llama-ds1}, the relative performance remains stable despite changes in BPW.
{\sys} achieves an average speedup of 2.05$\times$ (1.21$\times$) and 1.29$\times$ (1.74$\times$) over LRU and static baselines during the prefill (decode) stage, respectively.

In the QW-2 model, we reduce the cache rate as BPW increases from the default 4-bit to 16-bit to ensure the model fits within our 24\,GB GPU memory.
The decreased cache rate and increased memory footprint per expert limit {\sys}'s improvement.
For example, at INT4, {\sys} outperforms LRU by 2.06$\times$ in the prefill stage 
while the speedup drops to 1.105$\times$ at FP16.

\begin{figure}[t]
\begin{minipage}[t]{1\linewidth}
\centering
\includegraphics[width=1\linewidth]{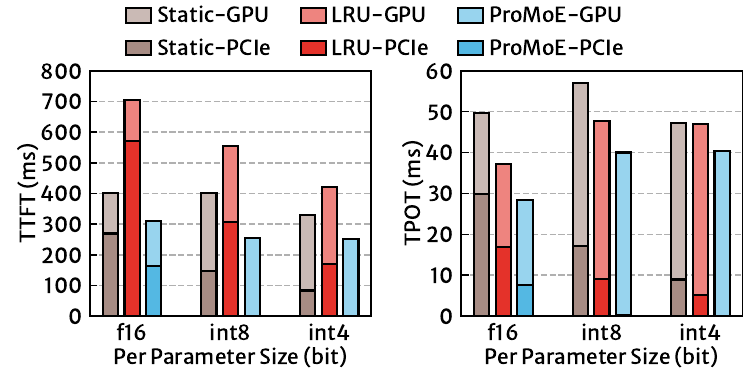}
\end{minipage} \begin{minipage}{1\linewidth}
\caption{\small{\emph{
The (a) TTFT and (b) TPOT of systems in transformers codebase with DS-1 model using different bits per weight.}}}
\label{fig:eval:sens-model-size-transformers-ds1}
\end{minipage}  \\[-10pt]
\end{figure}

We also conducted similar experiments in the transformers codebase using the DS-1 model, as illustrated in Figure~\ref{fig:eval:sens-model-size-transformers-ds1}.
The quantization was performed using the mainstream method GPTQ~\cite{gptq}.
In this scenario, {\sys} effectively overlaps the prefetching of experts with computations as BPW decreases.

\section{Related Work}
\label{sec:related}

\nospacestitle{Serving MoE-based LLMs with limited resources.}
Pre-gated MoE~\cite{pregatemoe} modifies the computation flow of MoE models by providing gate results for the subsequent layer from the previous layer directly, allowing accurate layer-wise prefetching by determining the required experts in advance.
SwapMoE~\cite{kong2024swapmoe} maintains a set of important experts in the GPU memory, using only these during inference to prevent offloading overhead.
In the background, it dynamically adjusts this set based on workload changes.
However, these systems alter the original MoE model computation, inevitably affecting model accuracy.
In contrast, {\sys} performs computations that are equivalent to the original model, accelerating the inference of MoE-based LLMs on edge devices without compromising accuracy.

Mixtral-offloading~\cite{mixtral_offloading} implements an LRU cache for the Mixtral MoE model 
and introduces a skip-based prediction method to support expert prefetching.
Brainstorm~\cite{brainstorm} designs a router abstraction to capture the dynamic aspects of models
and proposes speculative loading and execution based on static skewness statistics.
MoE-infinity~\cite{moe-infinity} develops an Expert Activation Tracing mechanism 
for sequence-level prediction to facilitate prefetching,
specifically designed for MoE-based encoder-decoder LLMs and 
aimed at throughput-oriented inference.
In contrast, {\sys} utilizes a learned predictor with high {\pmetric},
focusing on latency-oriented inference for edge devices.

\stitle{LLM serving on resource-constrained devices.}
Most modern frameworks~\cite{transformers,llamacpp,vllm,tgi,deepspeed} for serving LLMs provide basic offloading support that utilizes the CPU to handle parameters or computations, thereby reducing the GPU memory requirements.
FlexGen~\cite{sheng_flexgen_2023} aggregates memory and computation resources from the GPU, CPU, and disk.
It optimizes tensor storage and access patterns while also compressing weights and the attention cache.
These frameworks mainly target general LLMs and emphasize throughput-oriented inference with large batch sizes.
Model quantization~\cite{awq,gptq,hqq} and pruning~\cite{SparseGPT,wanda} are techniques to reduce the memory requirements of LLMs.
DejaVu~\cite{liu_deja_2023} takes advantage of contextual sparsity in LLMs to lower inference costs with minimal impact on model quality.
It employs a low-cost algorithm to predict input-dependent sparse subsets of attention heads and MLP parameters on-the-fly, which reduces the number of parameters needed during inference.
Building on DejaVu, PowerInfer~\cite{song_powerinfer_2023} utilizes the power-law distribution of neuron activations in LLMs, preloading frequently activated ``hot'' parameters onto the GPU while processing less active ``cold'' parameters on the CPU.
{\sys} is orthogonal to these techniques and can be integrated with them to further minimize memory usage and enhance inference speed.

\stitle{Generic LLM serving optimization.}
The rising demand for LLMs has prompted various system optimizations~\cite{specinfer,dao_flashattention_2022,specdecode,xia2023flashllm,medusa,tensorrt-llm} to improve their performance and efficiency.
vLLM~\cite{vllm} introduces PagedAttention, which manages the key-value cache for LLM serving and allows for sharing the cache across requests.
This improves batching efficiency and reduces memory fragmentation.
Orca~\cite{orca} proposes continuous and selective batching to optimize the performance of batched LLM serving.
While these systems focus on enhancing batched LLM serving in the cloud environment, {\sys} is designed specifically for low-latency, single-request LLM inference on edge devices.
 
\vspace{-3mm}
\section{Conclusion}
\vspace{-1mm}
This paper presents {\sys}, a proactive caching system that enhances expert offloading for MoE-based LLMs.
{\sys} leverages a learned predictor and carefully coordinates prefetching with inference. 
Our evaluation shows the efficacy and efficiency of {\sys}.

\bibliographystyle{ACM-Reference-Format}
\diffatomic{  
\bibliography{main}
}

\end{document}